\documentclass[seceq,jjapprn]{jjap}
\usepackage{graphicx}
\title{Aharonov-Bohm Oscillations in Photoluminescence
 from Charged Exciton in Quantum Tubes}
\author{Kohei ~\textsc{Tsumura}$^{1}$, Shintaro ~\textsc{Nomura}$^{1,2}$, 
 Pamela ~\textsc{Mohan}$^{3}$,  Junichi ~\textsc{Motohisa}$^{3}$,
 and Takashi ~\textsc{Fukui}$^{3}$ }
\inst{$^{1}$Institute of Physics, University of Tsukuba,
 Tennodai, Tsukuba, Ibaraki 305-8571, Japan
$^{2}$CREST-JST, 4-1-8 Hon-cho, Kawaguchi 332-0012, Japan \\
$^{3}$Research Center for Integrated Quantum Electronics,
 Hokkaido University, North 13 West 8, Sapporo 060-8628, Japan}

\abst{
The oscillation of photoluminescence peak energies is observed
in InAs quantum tubes depending on the magnetic flux through the tube.
The oscillation is shown
to be due to the Aharonov-Bohm effect of a charged exciton in a quantum tube.
No quadratic shift in photoluminescence peak energies is observed, which
is a characteristic feature of a thin quantum tube with a single channel
surrounding the magnetic flux through the tube.
}

\kword{nanotube, Aharonov-Bohm effect, magnetooptics}

\begin{document}
\maketitle

\sloppy

The Aharonov-Bohm (AB) effect is a sensitive probe of the phase difference
 of the wave function of a charged particle. 
By splitting the wave function of a charged particle in two paths and
 recombining them again, the phase difference acquired by the two paths
 gives the oscillation of the ground-state energy 
depending on the magnetic flux threading the area surrounded 
by the two paths. 
This quantum interference effect due to a gauge potential $\bf{A}(x)$ 
has raised considerable theoretical disputes. 
Direct evidence of the AB effect has been presented 
by observing the quantum interference pattern
in electron holograms~\cite{Tonomura86}.
The AB effect has been actively investigated by the transport measurements 
of metal-, super-, and semiconductor-mesoscopic rings. 
The AB effect in these systems is essentially understood
by considering the oscillation
of the single-particle energy of a charged particle 
in the limiting case where the dephasing due to impurities, 
phonons and electron-electron interaction is small.

The AB effect of an electron-hole composite system has recently been
proposed, where the oscillation of the ground state 
originates from the interference of electron and hole wave functions
traveling in two different paths~\cite{Chaplik95,Romer00}.
Not only the single-particle properties of an electron and a hole, 
but also the electron-hole Coulomb correlation play large roles in
the AB effect of an electron-hole composite system.
Experimental investigations have been performed to observe
the proposed AB effect of an electron-hole composite. 
The oscillation of the photoluminescence (PL) peak energy with 
magnetic fields was observed in a sample with a ring geometry,
but the oscillation was explained by a single particle energy of 
the final state electron after the recombination of 
a charged exciton~\cite{Bayer03,Korkusinski02}. 
Another approach is to measure the PL from type II quantum dots (QDs),
where electrons are confined inside and the holes are bound 
outside~\cite{Ribeiro04}. 
In this case, the detected oscillation of transition energy 
was found to be solely due to the oscillation of 
a single particle energy of the hole traveling around QDs.

The delicate nature of the AB effect of an electron-hole composite system
has been ascribed to the destructive interference 
induced by many different paths in a ring. 
In order to avoid this destructive interference, 
the ring must be sufficiently narrow, 
and the size of the ring must be smaller 
than the effective Bohr radius of an electron-hole composite. 
However, if the ring is infinitesimally narrow,
the amplitude of the AB oscillation is zero because the Coulomb
interaction of a one-dimensional exciton diverges~\cite{Banyai87,Ogawa91}. 
Therefore, an infinitesimally thin quantum tube
is an ideal system for observing the AB effect of an electron-hole composite system. 
A recently developed sophisticated crystal growth technique enables us 
to prepare a sample that satisfies the above conditions. 
By selective-area metal organic vapor phase epitaxy (SA-MOVPE), 
highly uniform arrays of InAs quantum tubes surrounded by InP barriers 
have been fabricated~\cite{Mohan05,Mohan06}. 
The interface of InP and InAs is atomically flat with  
InAs tubes with a thickness of several monolayers (MLs). 
In this study, we demonstrate the AB effect of an electron-hole composite
system originating from the phase difference between electron and hole 
wave functions induced by the oscillation of the PL peak with magnetic flux.

The samples studied are SA-MOVPE-grown InP/InAs/InP core multishell
nanowires (CMS-NWs) on InP (111)A substrates.
SA-MOVPE is one of MOVPE methods combined with electron beam (EB) lithography, 
which enables us to grow uniformly shaped structures
in arbitrary areas ~\cite{Mohan05}. 
A schematic structure and a high-resolution cross-sectional
scanning electron microscopy image of CMS-NW are shown in Fig. 1. 
The inner NW of radius 35 nm and the outermost shell are InP barriers. 
A few MLs of InAs are embedded between them so that  
a single type-I radial quantum well (QW) is formed. 
The variation in the thickness of the radial InAs QW in the vertical
 direction is controlled to be small. 
Failure to control the thickness would result in the bending of CMS-NWs. 
These CMS-NWs were excluded in PL measurement. 
Their typical length is $2 \mu\rm m$. 
Details of the growth method for and the characteristics of 
the sample can be found elsewhere~\cite{Mohan06}.

The sample was mounted on a cold-finger at 4.2 K in 
an optical exchange-gas He-cryostat equipped with a superconducting magnet.
Magneto-PL measurement was carried out in a magnetic field ($B$)
parallel to NWs up to 3 T by exciting the sample with a linearly polarized
light at 800 nm, which was focused onto a 80-$\mu\rm m$-diameter spot 
on the sample with a $\times 10$ objective lens with an N.A. of 0.26. 
The diameter of the laser spot was intentionally increased in order to reduce
the areal excitation density of the radial InAsQW, which was set
to be smaller than 20 $\rm W/cm^{2}$. 
CMS-NWs were grown perpendicular to the substrate, 
i.e., parallel to the incident light. 
The PL from the sample was collected by using the same objective lens, 
and analyzed using a Fourier-transform infrared (FTIR) spectrometer
equipped with a liquid-nitrogen-cooled HgCdTe detector. 
The density of CMS-NWs was controlled depending on its grown site 
on a single wafer by EB lithography. 
Their typical PL spectra at 0 T are shown in Fig. 2. 
The PL peaks labeled $P_{i}$ correspond to the peaks due to $N+i$-ML of 
InAs~\cite{Mohan06} with $N\approx0$. 
The variation in the PL peak energies of the CMS-NWs 
depending on the position of the sample is small, 
although the intensity of the peaks varies. 
This shows that CMS-NWs were grown uniformly in a large area, 
and the thickness of the InAs radial QW layer varies between CMS-NWs
at different positions. 
The thickness of the InAs radial QW layer is most probably the same 
in a single CMS-NW. 
The linewidths of the PL peaks are considered to be due to 
the distribution of the strain fields, 
but details are not known.

The PL peak energies of CMS-NWs obtained by fitting the PL spectra 
to a Gaussian-type spectral profile are shown in Fig. 3. 
The PL peak energies show an oscillation with a period of about 1 T.
The amplitude of the oscillation is about 1 meV for the three peaks. 
In contrast to the previous reports on InGaAs quantum rings~\cite{Bayer03}
or InP type-II QDs~\cite{Ribeiro04},
 no quadratic shift in the PL transition energies with magnetic field
is observed. This is exactly what is expected for a quantum ring~\cite{Chaplik95,Romer00}
or a quantum tube~\cite{Nomura07} with a single channel
surrounding the magnetic flux.

In order to analyze this observation, 
we model a quantum tube with radius $R$ in a magnetic field parallel 
to the axis of the tube ($\bf{e}_{z}$) within the effective-mass approximation 
by assuming an infinitesimally thin quantum tube. 
The electrons and  hole are confined in the radial QW, 
interacting by the Coulomb interaction $V_{e-h}(\bf{r})$. 

The Hamiltonian of the $X^{0}$ state is given by
\begin{equation}
{H}^{(X^{0})}=-{\hbar^{2} \over 2M^{X0}}
{\nabla}_{c.m.}^{2}+{\hbar^{2} \over 2{\mu}^*}
\left({{{{\nabla }_{rel} \over i}}^{}\rm +{\mit \Phi \over
 R{\phi}_{0}}{\bf e}_{\rm \varphi }}\right)^{2}+
{\it V}_{\it e-h}({{\bf r}_{\it rel}}),
\end{equation}
where $M^{X0}=m_{e}+m_{h}$, $z_{rel}=z^{(e)}-z^{(h)}$,
$z_{c.m.}^{\left({X^{0}}\right)}={m_{e} \over M^{X0}} z^{(e)}
+{m_{h} \over M^{X0}}z^{(h)}$,
$\varphi_{rel}=\varphi^{(e)}-\varphi^{(h)}$,
$\varphi_{c.m.}^{\left({X^{0}}\right)}={m_{e} \over M^{X0}} \varphi^{(e)}+{{m}_{h}
\over M^{X0}} \varphi^{(h)}$,
${\mu^{*}}^{-1}={m_{e}}^{-1}+{m_{h}}^{-1}$,
$m_{e}$, and $m_{h}$ are the effective masses of an electron and a hole,
respectively.
The energy of the center-of-mass motion is given by 
$E_{K, N_{c.m.}}^{(X^{0}, c.m.)}={\hbar^{2} \over 2{M}^{X0}}
\left[{K^{2}+{\left({{N_{c.m.} \over R}}\right)}^{2}}\right]$,
where $K$ is the wave number in the axis direction,
and $N_{c.m.}$ is the angular momentum of the center-of-mass motion.
We are interested in the optically active $X^{0}$ near the $\Gamma$ point
and the energy of the center-of-mass motion is set to zero, i.e.,
$K=0$ and $N_{c.m.}=0$. 

Since the InAs quantum tube layer embedded in InP barrier layers is 
compressibly strained, the heavy- and light-hole bands are strongly mixed. 
The effective mass of the upper hole band in the quantum tube 
along the circumference direction is given by 
$m_{h}=m_{0}/({\gamma}_{1}-\sqrt{3}{\gamma}_{2})$,
where $\gamma_{1}$ and $\gamma_{2}$ are Luttinger's parameters,
by assuming a relatively large splitting of the hole bands~\cite{Nishi94}.
By substituting $m_{e}=0.023 m_{0}$, $\gamma_{1}= 19.67$, and
$\gamma_{2}=8.37$, $m_{h}$ is estimated to be $0.19 m_{0}$. 
The effective exciton Bohr radius ($a_{B}^{*}=\epsilon \hbar^{2}/
\mu^{*} e^{2}$) and the effective Rydberg energy
($E_{Ry}^{*}=\mu^{*}{e}^{4}/{2\epsilon}^{2}\hbar^{2}$)
are respectively calculated to be $a_{B}^{*}=39$ nm,
and $E_{Ry}^{*}=1.2$ meV using $\epsilon= 15.15\epsilon_{0}$. 
The radius of the quantum tube $R=35 \rm nm$ corresponds 
to the magnetic flux quanta $\phi_{0} = 1.07$ T. 
With these material parameters, the lowest state energy of $X^{0}$ is 
calculated as a function of magnetic field. 
The derivatives with respect to $z_{rel}$ are expanded in real-space 
by the higher-order finite difference method~\cite{Chelikowsky94,Nomura04}. 
The ground state energy and wave function are obtained 
by the numerical diagonalization of the Hamiltonian matrix
as described in detail elsewhere~\cite{Nomura07}. 
The calculated result is shown in Fig. 4 (a). 
The lowest state energy of $X^{0}$ shows a sinusoidal oscillation
with a period of $\phi_{0}$. 
The maxima of the lowest state occur at half integer flux quanta, 
in contrast to the observation in Fig. 2, 
where the maxima of the PL peak energies occur at integer flux quanta. 
The small calculated amplitude of 0.0021 meV also excludes 
the experimental observation of the AB oscillation of a $X^{0}$ state.

This leads us to investigate the AB oscillation of a charged exciton ($X^{-}$) 
state ~\cite{Munschy74,Finkelstein95,Wojs95}. 
A $X^{-}$ state is a bound state of two electrons and a hole. 
As a result, the spatial extent of the wave function is larger for $X^{-}$
than for $X^{0}$. 
It is expected that the amplitude of AB oscillation is
larger in a $X^{-}$ state than in a $X^{0}$ state. 
Although our sample is not nominally undoped, our recent investigations
indicate that InP nanowires are actually $n$-type doped, presumably because
of the residual donor impurities in the source materials for the growth.
Moreover, in samples with a built-in electric field,
some electrons may be spatially separated from holes
after the dissociation of optically generated $X^{0}$. 
The observation of $X^{-}$ in undoped samples and mechanisms 
to provide additional free carriers have been reported ~\cite{Phillips96}.
 
The Hamiltonian of a $X^{-}$ state is separated into
the center-of-mass motion ($H_{c.m.}$) and the relative motion
with respect to a hole ($H_{rel}$)~\cite{Nomura07,Korkusinski02,Munschy74}
as given by $ H=H_{c.m.}+H_{rel}$,
$H_{c.m.}={\hbar^{2} \over 2M}{\left({{{\nabla}_{c.m.} \over i}+
{\Phi \over R{\phi}_{0}}{\bf e}_{\varphi}}\right)}^{2}$
and
\begin{equation}
H_{rel}=\sum_{j=1}^{2} \left[{{{\hbar^{2} \over 2{\mu}^{*}}
\left({{\nabla_{j} \over i}+{\Phi \over R\phi_{0}}{\bf e}_{\varphi}}\right)}^{2}
+V_{e-h}\left({{\bf r}_{j}}\right)}\right]\\
+V_{e-e}\left({{\bf r}_{1}-{\bf r}_{2}}\right)
-{{\hbar}^{2} \over {m_{h}}}\nabla_{1} \nabla_{2}
-{2 \sigma \over {1+2\sigma}} {\hbar^{2} \over 2 \mu^{*}}
{\left({{\Phi \over R \phi_{0}}}\right)}^{2},
\end{equation}
where ${\bf r}_{j}={\bf r}_{j}^{(e)}-{\bf r}^{(h)}$,
${\bf r}_{c.m.}={m_{e} \over M}\left({{\bf r}_{1}^{(e)}+{\bf r}_{1}^{(e)}}
\right)+{{m}_{h} \over M}{\bf r}^{(h)}$,
$M=2m_{e}+m_{h}$, $\sigma=m_{e}/m_{h}$,
$V_{e-h}({\bf r}_{j})$ and $V_{e-e}({\bf r}_{1}-{\bf r}_{2})$
are the attractive and repulsive Coulomb interaction terms, respectively.
The lowest energy of the singlet $X^{-}$ state is calculated,
which is the ground state in the magnetic field of interest.

Upon optical transition, an electron and a hole recombine, 
leaving an electron in the final state. 
The optical transition energy of $X^{-}$ is given by 
$E(X^{-})-E(e^{-}) = E_{rel} + E_{c.m.}-E_{sp}+E_{0}$, 
where $E_{rel}$, $E_{c.m.}$, $E_{sp}$, and $E_{0}$
are the energies of the electron-hole relative motion, 
the center-of-mass motion, the single electron in the final state, 
and the band-gap energy, respectively. 
With an increase in magnetic field, 
$E_{rel}$ and $E_{c.m.}$ have maxima at half integer flux quanta, 
while $-E_{sp}$ has minima at half integer flux quanta as shown in Fig. 4. 
By summing up three contributions, the transition energy shows an oscillatory structure
with a period of $\phi_{0}$ with minima located at half integer flux quanta, 
in agreement with the observation in Fig. 3. 
The $X^{-}$ and $X^{0}$ states are not resolved in Fig. 2 
because of the small energy separation $E(X^{0})-E(X^{-}) < 0.3$ meV. 
It should be noted that the oscillation of the optical
transition energy with a period of $\phi_{0}$ without diamagnetic shift
is only expected for an infinitesimally thin quantum tube~\cite{Nomura07}.
In a quantum ring with infinitesimally small width, the amplitude
of AB oscillation would be zero because of the divergence of the
Coulomb interaction, and in a quantum ring with a finite width,
the exciton ground state shows a quadratic shift with magnetic field~\cite{JSong01}.

Although our calculation is simple, the results capture
the essential features of the observed oscillation of PL energy. 
However, there are several points to be noted. 
The observed amplitude of the oscillation is not quantitatively explained 
by the calculation. 
The effective masses may be smaller than those of the bulk possibly 
because of the built-in strain field in the sample,
or the electron-density-dependent effective-mass renormalization~\cite{Asgari05}. 
The spectral linewidths in the PL spectra may also be accounted for 
by the distribution of strain fields. 
In the calculations, the thickness of the InAs radial QW layer, 
the hexagonal shape of the crosssection of the radial QW layer, 
and the distribution of the strain fields are not taken into account. 
Quantitative estimations of these effects are
beyond the scope of this study.
The Hamiltonian of a $X^{-}$ state becomes nonseparable
into the center-of-mass and the relative motions
if these effects are to be taken into account, 
which leads to an impractically large computational time. 

The possible localization of a hole does not change our arguments above. 
In this case, the calculated optical transition energy oscillates 
with a period of $\phi_{0}$ by substituting $m_{e}/m_{h}=\sigma=0$
in the limit  $m_{h} \rightarrow \infty$. 
The relative motion of an electron and a hole is not affected by the possible 
localization of a hole. 
The amplitude of the oscillation of optical transition energy
increases slightly with the
substitution of $E_{c.m.}=0$, $a_{B}^{*}=35$ nm, and $E_{Ry}^{*}=1.4$ meV 
by setting $\sigma=0$. 
It is difficult to completely exclude the possibility
of the localization of a hole.
However, because PL efficiency would decrease markedly 
in the presence of defects or impurities in our very thin quantum tube structure
with a small diameter
and only samples
with a high PL efficiency and a high homogeneity are selected for measurements,
we exclude the possibility of the localization of a hole.

As shown in Fig. 4, the oscillation amplitude of the single particle energy
of the electron in the final state is larger than that of the
energy of the electron-hole relative motion.
However, the oscillation of the energy of the electron-hole relative motion
is expected to be dominant for samples with a smaller diameter
because the single particle
energy increases with $1/R^{2}$, while the energy of the electron-hole relative motion follows $\rm exp(\it -CR)$, where $C$ is a constant.~\cite{Nomura07}
The change in the optical transition energy of  $X^{-}$
from minima to maxima at half integer flux quanta is expected to be observed
with the reduction in the diameter of a quantum tube
as a qualitative evidence of the excitonic AB effect. 

In conclusion, the oscillation of PL peak energies is observed depending 
on $\Phi$ in a CMS-NW sample. 
The maxima of PL peak energies are found to occur at integer flux quanta. 
This oscillatory structure with a period of $\phi_{0}$ is explained 
by the AB effect of the $X^{-}$ exciton state.
No quadratic shift in PL transition energies is observed, which
is explained by results of calculation for an infinitesimally thin quantum tube.
This is the main qualitative difference between our results
and previous results.~\cite{Bayer03,Ribeiro04}
Our results demonstrate that it is feasible to study the electronic properties
of artificially designed nanostructures with atomic precision.

\acknowledgements{
Part of this work was supported by a Grant-in-Aid for Scientific Research
 by the Japan Society for the Promotion of Science,
 and the Nano-Science Special Project of the University of Tsukuba. 
}

\begin{halffigure}
\caption{
(a) Schematic illustration of sample structure. 
(b) Scanning electron microscopy image of quantum tubes.
(c) Schematic cross-sectional image of quantum tube. 
(d) High resolution cross-sectional scanning electron microscopy image
of a quantum tube sample observed after anisotropic dry etching
and stain etching. }
\label{sample}
\end{halffigure}

\begin{halffigure}
\caption{
PL spectrum of CMS-NWs at $B=0$ T.
(inset) PL spectra measured at 10 different positions on single wafer.}
\label{microPL}
\end{halffigure}

\begin{halffigure}
\caption{
Magnetic field dependence of PL peak energies, (i) $P_{1}$,
(ii) $P_{2}$, and (iii) $P_{3}$, of quantum tube sample.}
\label{PLpeaks}
\end{halffigure}

\begin{halffigure}
\caption{
(a) Calculated optical transition energy of $X^{0}$ state.
The origin of the vertical axis is the band gap energy.
(b) Calculated magnetic field dependence of (i) $E_{c.m.}$, (ii) -$E_{sp}$,
and (iii) $E_{rel}$ and (iv) $E_{rel}+E_{c.m.}-E_{sp}$ of $X^{-}$ state. 
}
\label{Calc}
\end{halffigure}

\makefigurecaptions

\includegraphics[width=15.0cm]{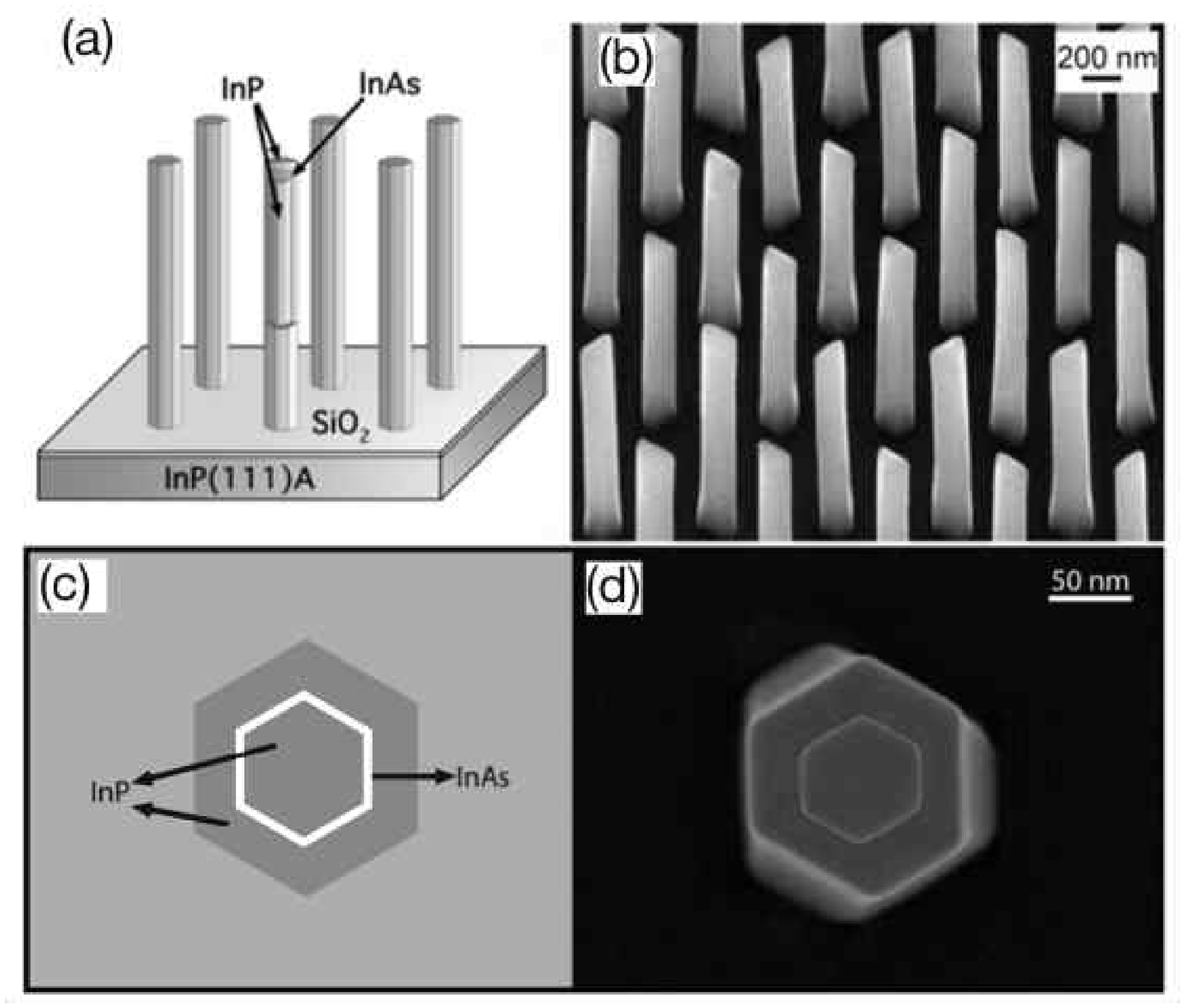}
\newpage

\includegraphics[width=15.0cm]{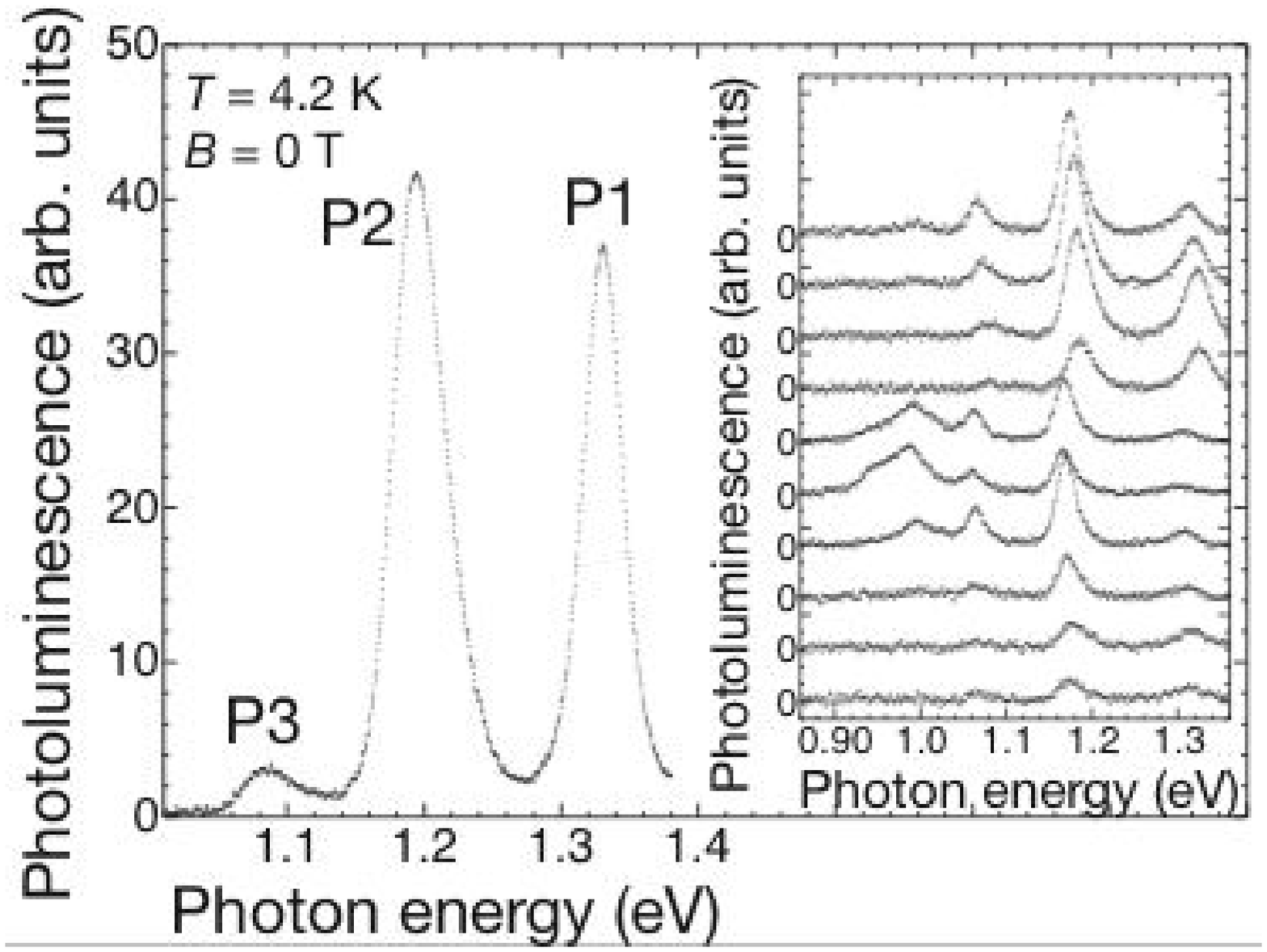}

\includegraphics[width=15.0cm]{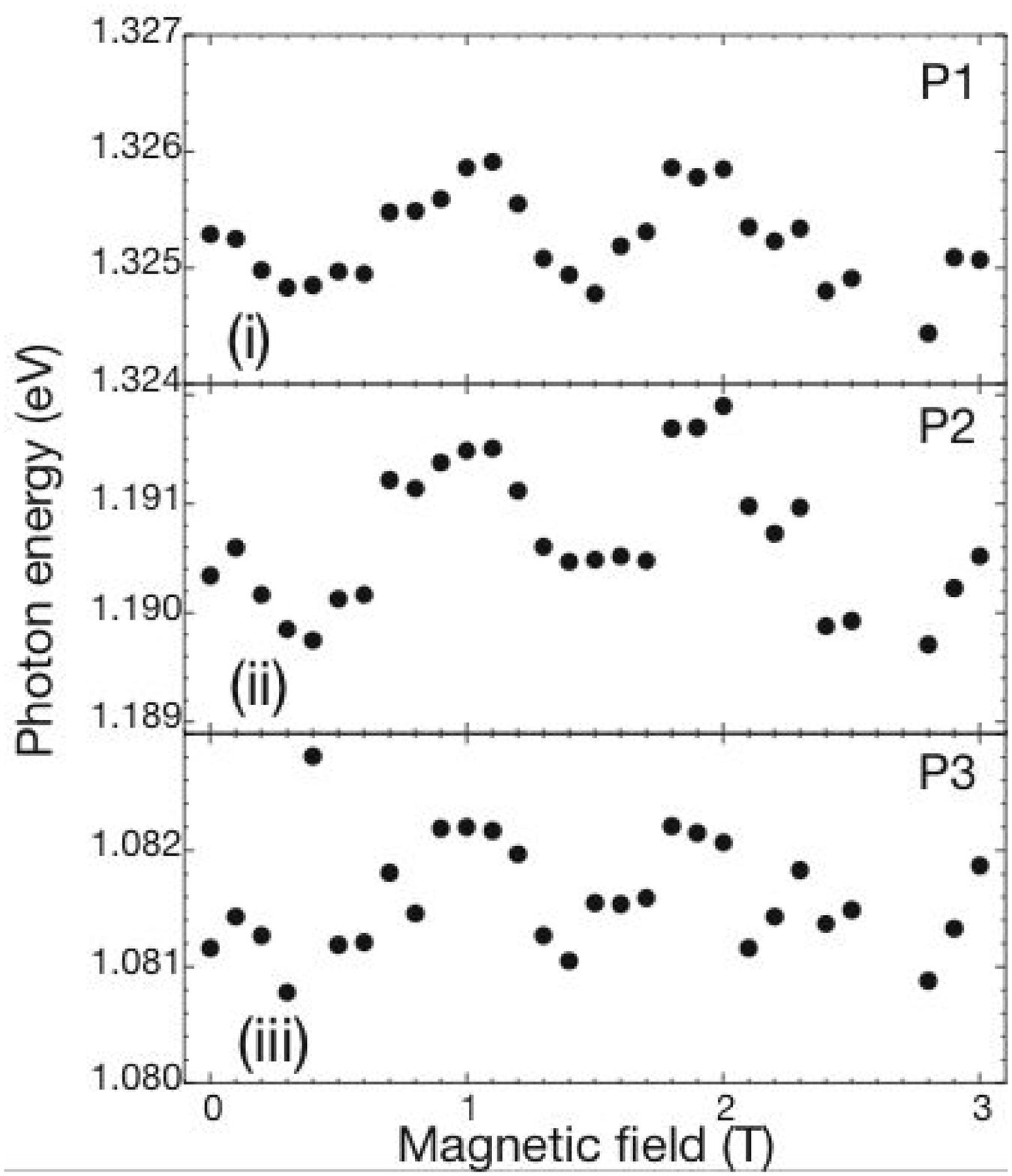}

\includegraphics[width=15.0cm]{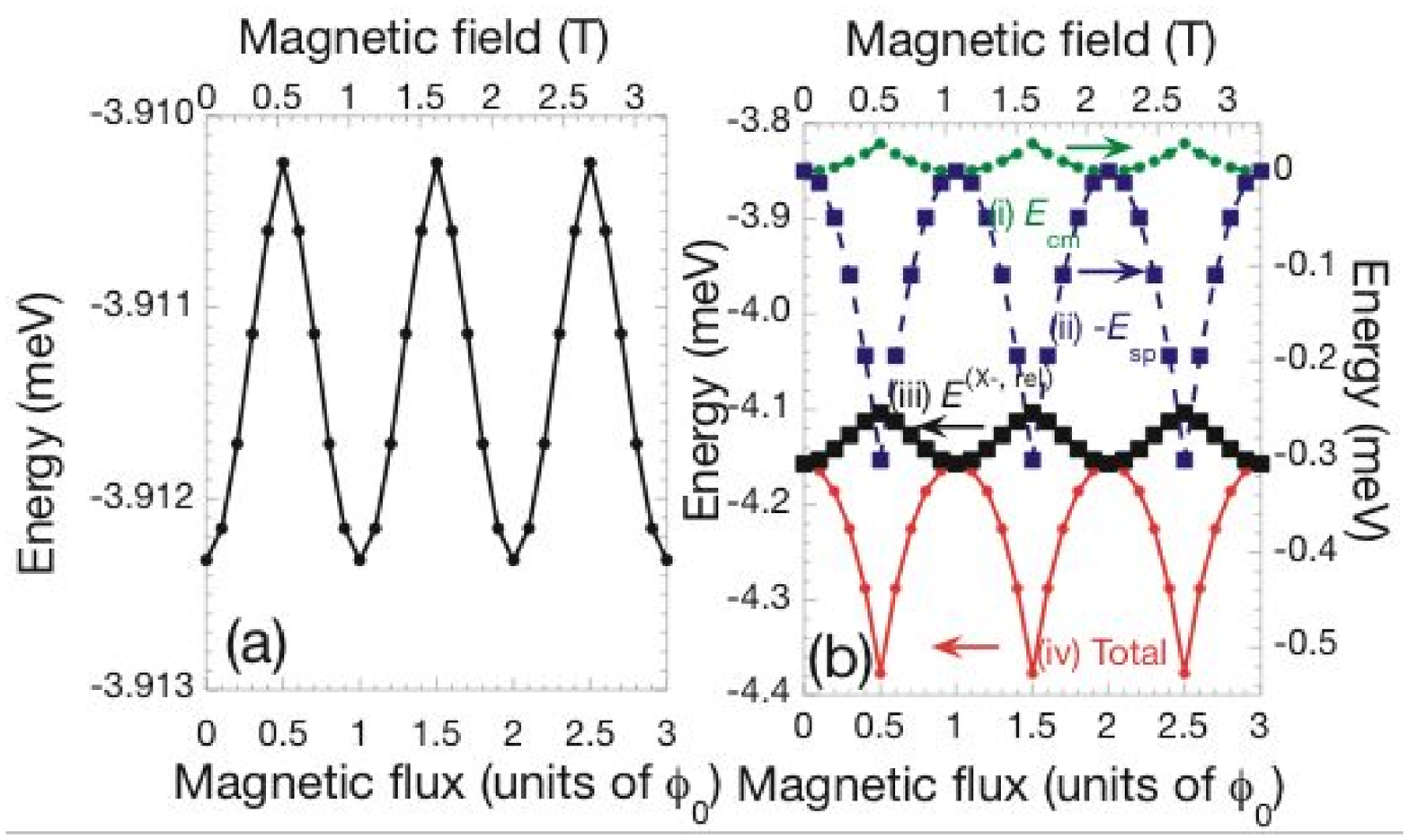}

\end{document}